 \documentstyle[12pt]{article}
 \def \bfgr #1{ \mbox {{\boldmath $#1$}}}
\begin{document}

\begin{center}

{\Large \bf
 Parametrization of Realistic Bethe-Salpeter Amplitude
 for  the Deuteron
 }

\vskip 5mm 

{A.Yu.  Umnikov}
\footnote{INFN Postdoctoral Fellow}

 \vskip 5mm 

{\em Department of Physics, University of Perugia, and
INFN, Sezione di Perugia,\\
via A. Pascoli, Perugia, I-06100, Italy.}

\end{center}
\vskip 5mm

\begin{abstract} 
The parametrization of the realistic
Bethe-Salpeter amplitude for the
deuteron is given. Eight components of the amplitude in the
Euclidean space
are presented as an analytical fit to the numerical
solution of the Bethe-Salpeter equation in the ladder approximation.
An applicability
of the parametrization to the   
observables of the deuteron is briefly discussed.
\end{abstract}

\section{Motivations}

Since the first solution of the realistic Bethe-Salpeter (BS) 
equation for the deuteron~\cite{tjond}, the BS-amplitudes
have been applied to describe  various processes
with the deuterons~\cite{tjond,tjon2,ukkk,kuk,ukk}. The obvious advantages 
of the  approaches based on the BS formalism are the explicit
covariance and connection to the covariant dynamical (field) theory.
In spite of this, practical use of the BS amplitudes is not as popular
as of nonrelativistic wave functions~\cite{paris,bonn}.
A more complicated physical interpretation and  technical   complexity
of the approaches based on the BS amplitudes are the main reasons for that.

 In a series of recent papers
 it has been argued that a new intuition can be developed in working with the 
 BS amplitudes~\cite{ukkk,kuk,ukk}. It has been shown that 
 calculations of   many observables for the deuteron
 are reduced to calculations similar to those in  field theory.
 It   has also been stressed that the usage of the BS amplitude for the
 deuteron in the Dirac matrix basis can be more convenient than in the spinor 
 basis. In this case  computations are formalized
 enough to extensively apply  analytical 
 computing software, such as the Mathematica~\cite{wolf}, 
 to calculate  the matrix elements of  observables in terms of the 
 components of the BS amplitude.
 However, the promotion of a new technique should assume that  all basic ingredients
 of the calculation are
 available to a potential user. The BS formalism still lacks this feature,
 since there is no simple parametrization available for the realistic 
 deuteron amplitude, such as is available for nonrelativistic wave functions~\cite{bonn,paris1}
 or relativistic wave functions of the spectator equation~\cite{buckgross}.

 This letter  presents the analytical parametrization of the
 Bethe-Salpeter amplitude for the deuteron. 
 The  parameters are fixed
 by  fitting  to the recent numerical solution of the homogeneous BS equation
 using the Dirac matrix basis~\cite{uk,ukk}. The  one-boson exchange potential 
 from ref.~\cite{tjond,tjonnn}
 was used with a minor adjustment of its parameters~\cite{ukkk}, so in this sense
 the solution does not contain new physics (or different physics) then
 the pioneer paper.

\section{Definitions and kinematics}

The realistic BS amplitude of the deuteron, $\Phi $, can be obtained as
a solution to the homogeneous BS equation with the effective 
one-boson-exchange kernel~\cite{tjond,ukkk}:
\begin{eqnarray}
\Phi(p,P_D) = { i} S(p_1,p_2)\sum_{B} \int\frac{d^4p'}{(2\pi)^4}
  \frac{g^2_B\Gamma^{(1)}\otimes\Gamma^{(2)}}{(p-p')^2-\mu^2_B+i\epsilon}
  \Phi(p',P_D),
\label{bs}
\end{eqnarray}
where $P_D$ is the deuteron momentum,
$\mu_B$ is the mass of the meson $B$, $\Gamma_B^{(k)}$ is the meson-nucleon
vertex, corresponding to the meson $B$ and connected to the $k$-th nucleon. The tensor
product in the r.h.s. of eq. (\ref{bs}) is for a possible interconnection
of the quantum numbers in two vertices 
(such as in the case of the vector  or isovector mesons). The two-nucleon
propagator, $S(p_1,p_2)$, is defined as:
\begin{eqnarray}
S(p_1,p_2) \equiv \frac{\hat p_1 +m}{p^2_1-m^2+i\epsilon}
\cdot \frac{\hat p_2 +m}{p^2_2-m^2+i\epsilon}=
(\hat p_1 +m)(\hat p_2 +m)D(p,P_D), \label{prop}\\[2mm]
\quad p_{1,2} = \frac{P_D}{2}\pm p,
\label{p12}
\end{eqnarray}
where $m$ is the nucleon mass, $p_{1,2}$ are nucleon momenta,  $\hat p_k = \gamma_\mu p^\mu_k$
and $D(p,P_D)$ is ``scalar two particle propagator''.

It is convenient 
to use the amplitude, $\Psi$, conjugated with respect 
to one of the nucleon lines~\cite{uk}:
\begin{eqnarray}
\Phi = \Psi \gamma^c \quad {\rm or} \quad \Psi = -\Phi \gamma^c,
\label{conj}
\end{eqnarray}
where $\gamma^c = \gamma_3\gamma_1$.
Then the amplitude $\Psi $ can be decomposed in terms of a
complete set of 
the Dirac matrices, their bilinear combinations and the
 $4\times 4$-identity matrix, $\hat{\bfgr 1}$, with sixteen 
 components being the coefficients of decomposition:
\begin{eqnarray}
\Psi = \hat{\bfgr 1}\psi_2+\gamma_5\psi_p+\gamma_\mu \psi^\mu_v
+\gamma_5\gamma_\mu\psi^\mu_a+\sigma_{\mu\nu}\psi^{\mu\nu}_t.
\label{dirac}
\end{eqnarray}

In the deuteron rest frame the notions are used:
\begin{eqnarray}
&&\psi_v^\mu \equiv  (\psi_v^0, {\bfgr \psi}_v) , \quad
\psi_a^\mu \equiv (\psi_a^0, {\bfgr \psi}_a) ,
\label{restnot}\\[2mm]
&&\psi_t^{0i} \equiv  {\bfgr \psi}_t^0,\quad \psi_t^{ij} \equiv  
\varepsilon^{ijk}\psi_t^k \quad \left[ \psi^k_t \equiv {\bfgr \psi}_t\right],
\label{restnot1}
\end{eqnarray}
where $i,j,k = 1,2,3$ and other tensor components of $\psi_t^{\mu\nu}$ are
equal to zero.

In order to separate the amplitude with the
deuteron's quantum numbers,  a partial wave
decomposition 
of the  four vector 
and four scalar functions, (\ref{dirac})-(\ref{restnot}), is performed:
\begin{eqnarray}
\psi=&& \sum_{JM} \psi(p_0,|{\bfgr p}|; J ){\sf Y}_{JM}(\Omega_p)
\label{partial}\\[2mm]
{\bfgr \psi} =&& \sum_{JM}\left \{
 \psi(p_0,|{\bfgr p}|; J-1J ){\bfgr Y}_{JM}^{J-1}(\Omega_p)\right. \nonumber\\
 &&+\left.
  \psi(p_0,|{\bfgr p}|; J J ){\bfgr Y}_{JM}^J(\Omega_p) +
   \psi(p_0,|{\bfgr p}|; J+1 J ){\bfgr Y}_{JM}^{J+1}(\Omega_p)
   \right\},
\label{partial1}
\end{eqnarray}
where ${\sf Y_{JM}}$ and ${\bfgr Y}_{JM}^J$ are the 
spherical harmonics and vector spherical harmonics respectively.

Fixing the total momentum $J=1$ and separating the components with 
positive parity, the deuteron's amplitude components read as:
\begin{eqnarray}
\psi_p(1 ),\; \psi_a^0(1 ),\; \psi_v(11 ), \;
 \psi_a(01 ), \; \psi_a(21 ), \;
\psi_t^0(11 ), \; \psi_t(01 ), \; \psi_t(21 ).
\label{deutcomp}
\end{eqnarray}
The components $\psi_a^0(1)$ and $\psi_t^0(11)$ are odd functions of $p_0$ and
all others are even.  

The BS equation with the realistic
one-boson exchange potential
is solved using the Wick rotation, which corresponds to
replace $p_0 \to ip_4$ and $\psi_a^0(1) \to i\psi_a^0(1)$~\cite{uk}. 
This procedure removes singularities from the exchange meson propagators
in  eq. (\ref{bs}) and from the
scalar propagator, $D$, which in the deuteron rest frame takes the form:
\begin{eqnarray}
D(p,P_D)^{-1}=D(p_4,|{\bfgr p}|)^{-1} = 
[m^2+{\bfgr p}^2+p_4^2-\frac14 M_D^2]^2 +p_4^2M_D^2,
\label{propW}
\end{eqnarray}
where $M_D = 2m+\epsilon_D$ is the deuteron mass.

After the Wick rotation, the components of the 
deuteron amplitude are {\em computed} along the imaginary axe in the 
complex $p_0$-plane.
Since  the inverse Wick rotation  of the 
{\em numerically } known amplitude is an ill-defined operation,
the parametrization for the components is obtained
in the Wick rotated case. 
The possibility to analytically continue those amplitudes into 
a physical region is discussed in  Section~4.

\section{The parametrization}

The parametrization of all components has the form (index $J=1$ is omitted):
\begin{eqnarray}
\psi (p_4, |{\bfgr p}|; L) = f(p_4, |{\bfgr p}|;L)
Exp\left \{
 g( |{\bfgr p}|; L) p_4^2
\right \} D(p_4,|{\bfgr p}|)m^4,
\label{form}
\end{eqnarray}
where $L = J, J\pm 1$, depending on the quantum numbers of the component and
functions $f$ and $g$ are given by:
\begin{eqnarray}
 f(p_4,|{\bfgr p}|; L) = \sum_{i=0}^{N_f} A_i(p_4)
  \frac{|{\bfgr p}|^L }{\alpha_i^2 + {\bfgr p}^2}, \quad
g(|{\bfgr p}|; L) = \sum_{i=1}^{N_g} B_i
  \frac{|{\bfgr p}|^L }{\beta_i^2 + {\bfgr p}^2}.
\label{forms2}
\end{eqnarray}
The form of parametrization (\ref{forms2}) and the scale   of the
parameters $\alpha_i$ are prompted by the previous works with 
parametrizations of the wave functions~\cite{bonn,paris1,buckgross}:
\begin{eqnarray}
&& \alpha_0 =  \mu_0/\sqrt{2}, \quad
  \alpha_i = i \mu_0, \quad i=1\ldots N_f, \nonumber\\
&& \beta_i = i \mu_1, \quad i=1\ldots N_g, \nonumber\\
&&\mu_0 = 0.139~{\rm GeV}, \quad \mu_1 = 2\mu_0,\nonumber\\
&& m = 0.939~{\rm GeV}, \quad M_D = 2m+\epsilon_D,\quad \epsilon_D = -2.2246~{\rm MeV}.
\label{par1}
\end{eqnarray}
The number $N_f$ is equal to 11 for all components, whereas
$N_g$ differs for different components.
The
$p_4$-dependence of the coefficients $A_i(p_4)$ is given by:
\begin{eqnarray}
&&A_i(p_4) =A_i + p_4^2 A_i' ,\phantom{p_090} \quad {\rm for} 
\quad \psi_p(1),\psi_v(1),\psi_a(0),\psi_a(2),\psi_t(0),\psi_t(2);\nonumber \\
&&A_i(p_4) =p_4(A_i + p_4^2 A_i' ), \quad {\rm for}\quad \psi_a^0(1),\psi_t^0(1). 
\label{A}
\end{eqnarray}

Coefficients $A_i$, $A_i'$ and $B_i$ for all components 
are given in   Appendix A, Tables~\ref{t1}-\ref{t8}.
 The presented parametrization contains a seemingly large number of parameters,
 27 to 29 per every of eight components plus three parameters common for all
 of them, including the nucleon mass, $m$, deuteron binding energy, $\epsilon_D$,
 and an
 additional mass scale parameter, $\mu_0$. This looks rather unusual for such type of
 parametrizations.  The parametrizations of  wave functions~\cite{bonn,paris1,buckgross},
 for instance,
 contain only $n \sim 10$ parameters per component.
 The reason for this difference is that the components of the BS amplitude
 depend  upon two independent variables, the relative momentum, $p$, and ``relative
 energy'', $p_4$. It is clear now, that the presented parametrization contain 
 quite a modest number of parameters;  it is not even $n^2$ compared to the 
  one dimensional fit of the wave functions.

\section{The applicability  of the parametrization}

The  parametrization 
(\ref{form})-(\ref{A}) is obtained by fitting the numerical solution
to the BS equation, using the least-squares procedure~\cite{wolf}.
 The domain of   validity of the parametrization
in 
relative momentum is $0<|{\bfgr p}| < 3$~GeV, means that 
the solution of the BS equation was fitted up to this point.
 The domain of validity 
in relative energy
 $ p_4$ (which is actually $ip_0$) is defined as follows. First,
 the
 singular structure of the BS amplitude in the Minkowski space
 is governed by the singularities of the propagator $D(P_D,p)$,
 eqs. (\ref{prop}) and (\ref{propW}), where the closest nucleon pole
 is  most important for the physical applications. Thus,
 the parametrization is valid at least up to 
  $p_4 \sim M_D/2 - \sqrt{m^2 + |{\bfgr p}|^2}$, corresponding to
 the nucleon pole at given $|{\bfgr p}|$. Second, 
 the parametrization allows for the integration in the matrix elements
  over
  $p_4$ with infinite limits,$(-\infty,+\infty)$, in the 
  Euclidean space.

The starting point for calculating any quantity with
the BS amplitude is the {\em relativistic impulse approximation}.
In many   cases the relativistic impulse approximation
is presented by the Feynman
 ``triangle
diagram'' with zero transfer of the momentum~\cite{uk,ukkk,kuk,ukk},
$q=0$ (Fig.~\ref{tre}):
\begin{eqnarray}
\langle \hat O \rangle = \int \frac{d^4p}{(2\pi)^4}{\sf Tr}\left\{
\bar\Psi(p_0,{\bfgr p})\hat O  \Psi(p_0,{\bfgr p}) (\hat p_2-m)
\right \},
 \label{me}
\end{eqnarray}
where $\bar\Psi = \gamma_0 \Psi^{\dagger}\gamma_0 $.
Two important examples are the matrix elements of the 
vector and axial currents, $\hat O = \gamma_\mu,\gamma_5\gamma_\mu$.

The matrix element $\langle \gamma_0 \rangle$,
the vector charge, is used to normalize the BS amplitude:
\begin{eqnarray}
1 = &\frac{1}{2M_D}& \int \frac{d^4p}{(2\pi)^4}{\sf Tr}\left\{
\bar \Psi(p_0,{\bfgr p}) \gamma_0    \Psi(p_0,{\bfgr p}) (\hat p_2-m)
\right \}  \label{vec}  \\
 =   &-\frac{1}{M_D}&  \int \frac{dp_4d{|{\bfgr p}|}{\bfgr p}^2}{(2\pi)^4}
  \left \{ \phantom{I^2}\!\!\!\!\!\!\nonumber  \right.\\
       &&   \quad\quad  -8 m \left(  \psi_{a}(0) \psi_{t}(0) + 
       \psi_{a}(2) \psi_{t}(2) \right) \nonumber \\ 
       &&  \quad \quad  +\frac{4p}{\sqrt 3}   \left( {{-2 \psi_{p}(1) \psi_{t}(0)}  
        } + 2 {\sqrt{{2 }}} \psi_{p}(1) 
      \psi_{t}(2) \right. \nonumber \\ 
      &&\quad\quad \quad\quad   \quad\left. + 
       {\sqrt{{2 }}} \psi_{a}(0) 
      \psi_{v}(1) + 
     {{  \psi_{a}(2) \psi_{v}(1)} }\
      \right) \nonumber  \\
            &&\!\!\!\!\!\!\!\!+(M_d - 2p_4)  
   \left(   {{\psi_{a}^0(1)}^2} +   {{\psi_{a}(0)}^2} + 
       {{\psi_{a}(2)}^2} +   {{\psi_{p}(1)}^2} \right.\nonumber  \\ 
     &&  \quad\quad \quad\quad \quad\left.  +   {{\psi_{v}(1)}^2} + 
      4 {{\psi_{t}(0)}^2} + 4 {{\psi_{t}(2)}^2} + 
   \left.  4 {{\psi_{t}^0(1)}^2} \right) \right \}. \label{no}
\end{eqnarray}
The components in eq.~(\ref{no}) are parametrized by eqs.~(\ref{form})-(\ref{A}). (Note that factor
$2\pi$ from integration over angle $\phi$ is absorbed by the parametrization.)
Integrating over $p_4$ in eq.~(\ref{no}) but keeping $|{\bfgr p}|$-dependence,
  one gets the charge density in the momentum space,
analogous to the square of the deuteron wave function in a nonrelativistic approach.
This charge density is shown in Fig.~\ref{den} together with the density of the 3-rd component
of the  axial current (omitting terms, vanishing after integration over  $\theta$):
\begin{eqnarray}
\langle \gamma_5\gamma_3  \rangle = &\frac{1}{2M_D}& \int \frac{d^4p}{(2\pi)^4}{\sf Tr}\left\{
\bar \Psi(p_0,{\bfgr p}) \gamma_5\gamma_3    \Psi(p_0,{\bfgr p}) (\hat p_2-m)
\right \}  \label{ax}  \\
 =   &\frac{1}{M_D}&  \int \frac{dp_4d{|{\bfgr p}|}{\bfgr p}^2}{(2\pi)^4}
  \left \{ \phantom{I^2}\!\!\!\!\!\!\nonumber  \right.\\
       &&   \quad\quad\quad  2 m \left(  4\psi_{a}(0) \psi_{t}(0) - 
       2\psi_{a}(2) \psi_{t}(2) \nonumber \right.\\
       && \quad\quad \quad\quad \quad\left.
      - 2\sqrt{2}\psi_{a}^0(1) \psi_{t}^0(1) 
      + \sqrt{2}\psi_{p}(1) \psi_{v}(1) 
       \right) \nonumber \\ 
       &&  \quad \quad  +\frac{2p}{\sqrt 3}   \left( {{4 \psi_{p}(1) \psi_{t}(0)}  
        } + 2 {\sqrt{{2 }}} \psi_{p}(1) 
      \psi_{t}(2) \right. \nonumber \\ 
      &&\quad\quad \quad\quad   \quad\left. -2 
       {\sqrt{{2 }}} \psi_{a}(0) 
      \psi_{v}(1) + 
     {{  \psi_{a}(2) \psi_{v}(1)} }\
      \right) \nonumber  \\
            &&\!\!\!\!\!\!\!\!+(M_d - 2p_4)  
   \left(   -{{\psi_{a}(0)}^2} + 
      \frac{1}{2} {{\psi_{a}(2)}^2} -\frac{1}{2}   {{\psi_{v}(1)}^2}\right.\nonumber  \\ 
     &&  \quad\quad \quad\quad \quad\left.   - 
      4 {{\psi_{t}(0)}^2} + 2 {{\psi_{t}(2)}^2} - 
   \left. 2 {{\psi_{t}^0(1)}^2} \phantom{\frac{1}{2}}\!\!\!\right) \right \}. \label{nox}
\end{eqnarray}

 The quality of the parametrization has been checked by a comparison of the charge and axial
 densities, as well as their integrals, computed using the parametrization and 
 original numerical components.
  The original amplitude is normalized by (\ref{vec}) and
 is exactly  equal to 1.0, whereas the parametrization gives the normalization equal 
 to 0.9997. 
 This is not a trivial result, since all components were fitted independently. One can
 use this number for the ``renormalization'' of observables. The original 
 amplitude gives an axial charge value of 0.9215, while the parametrization
 yields the same. The error  of the parametrization describing the densities
 is $\sim 0.01\%$ at small $|{\bfgr p}|$ to $\sim 1-2 \%$ at  $|{\bfgr p}|\sim $ 1-3~GeV.

Finally, the issue of an ``inverse Wick rotation'' should be addressed. The analysis
of the singular structure of the ``triangle graph'' and behavior of the BS amplitude
result in the conclusion that, perhaps, the presented parametrization can be used for
an analytical continuation, $p_4 \to -ip_0$,
 of the BS amplitude up 
 to the closest nucleon pole
of  $p_0 = M_D/2 - \sqrt{m^2 + |{\bfgr p}|^2}$. However, such a procedure
works well  (with accuracy $\sim 10\%$) only up to $|{\bfgr p}| \sim m$.  The accuracy was
estimated by calculating the  vector and axial
densities for the processes with the 
second nucleon on mass-shell.
If it is used further, the procedure gives 
an accuracy of $\sim 50\%$ at $|{\bfgr p}| = 1.5$~GeV
and it should not be used beyond this point.

\section{Summary}
 The   parametrization of the realistic 
Bethe-Salpeter amplitude for the deuteron has been presented. 
All eight components of the amplitude are given
in the Wick rotated case in the form of analytical functions.
Simple examples of the use of the parametrization are presented
and the applicability
 domain  is discussed.

\section{Acknowledgements}
 It is a pleasure to acknowledge
 stimulating conversations with L. Kaptari and F. Khanna.
 I would like to thank D. White for reading the manuscript
  and comments.

\newpage

\section*{Appendix A.  Tables of parameters}

{
\begin{table}[h]
\caption{Parameters for the $\psi_p(1)$ component}

\begin{tabular}{|c|c|r|r|r|}
\hline
     $L$ & $i$ & $A_i$& $A_i'$ &  $B_i$   \\
     \hline
     \hline
1&0&  -0.0108242476&        -34.3742135    &     --  \\
&1&    0.0576888290    &     119.943051     &   0\\
&2&     -3.00806368   &     -1257.32744    &   0\\
&3&      52.5918636    &     13865.0551     &   -222.089494  \\
&4&     -549.793133   &     -101125.416     &    890.102891  \\
&5&      3325.75711    &     472699.731     &   -1210.29749 \\
&6&     -11944.6774   &     -1417223.72    &     548.300818 \\
&7&      26383.0580    &     2743193.30    &     --  \\
&8&     -34689.7448   &     -3406750.03     &     --  \\
&9&      24977.5913    &     2621309.83  &     --  \\
&10&    -8159.59409   &     -1137897.83     &     --  \\
&11&     607.051002   &      213102.794      &     --  \\
\hline
\end{tabular}
\label{t1}
\end{table}}

{
 \begin{table}[h]
\caption{Parameters for the $\psi_a^0(1)$ component}

\begin{tabular}{|c|c|r|r|r|}
\hline
     $L$ & $i$ & $A_i$& $A_i'$ &  $B_i$   \\
     \hline
     \hline
1 &0 & 0.113315679 &    15.2791178 &    -- \\
 &1 & -0.565717405 &   -42.6551270  &   0\\
 &2 &   12.3795242 &    292.530922  &   0\\
 &3 &  -211.521767  &  -2058.70255  &   0\\
 &4 &   2025.71260  &   8992.57139  &   -48.3735989\\
 &5 &  -11769.4095  &  -20393.3149 &     91.9530671 \\
 &6 &   43037.5609 &    10486.7684  &   -44.7438523\\
 &7 &  -101273.551  &   55592.5995  &    -- \\
 &8 &   151268.732  &  -141733.721  &    -- \\
 &9 &  -136889.618  &   152338.654  &    -- \\
 &10 &  67943.7391 &   -80778.8808 &    -- \\
 &11 & -14143.4907  &   17289.2804  &    -- \\
\hline
\end{tabular}
\label{t2}
\end{table}}

\newpage

{
 \begin{table}[h]
\caption{Parameters for the $\psi_v(1)$ component}

\begin{tabular}{|c|c|r|r|r|}
\hline
     $L$ & $i$ & $A_i$& $A_i'$ &  $B_i$   \\
     \hline
     \hline
1 &0 &0.00545480918  &   0.452976149  &    -- \\
 &1  &-0.0413469671  &    3.58498830   &  0\\
 &2  & -0.626170875  &    49.3118036   &  0\\
 &3  &  -12.5904324   &  -435.917083    &  0\\
 & 4 &   144.575745   &   3114.24317   &   6.14953994 \\
 & 5 &  -1128.73423   &  -14069.8986   &  -25.2161192  \\
 & 6 &   4901.27381   &   41906.4275    &  19.9844772  \\
 & 7 &  -13706.4378   &  -85174.0821  &    -- \\
 & 8 &   24829.8901  &    114516.487   &    -- \\
 & 9 &  -26877.9006   &  -95556.5511  &    -- \\
 &10 &   15516.3123   &   44352.4015 &    -- \\
 & 11 & -3665.63223   &  -8706.31030  &    -- \\
\hline
\end{tabular}
\label{t3}
\end{table}}

{
\begin{table}[h]
\caption{Parameters for the $\psi_a(0)$ component}

\begin{tabular}{|c|c|r|r|r|}
\hline
     $L$ & $i$ & $A_i$& $A_i'$ &  $B_i$   \\
     \hline
     \hline
0 &0 &  -0.00198690442  &    -2.91350897 &    -- \\
 &1  &   0.01915782253  &   12.2001594   &  -0.853805793 \\
 &2  &   -0.154455300   &   -214.362824  &     17.7585282 \\
 &3  &     25.2282913   &    3289.28876   &   -89.3945864 \\
 &4  &    -361.376166  &    -30582.8134   &    155.578411 \\
 & 5 &     2946.98165   &    166962.671   &   -92.0411385\\
 & 6 &    -13477.6590   &   -553839.568 &    -- \\
 & 7 &     37613.2426   &    1152552.96   &    -- \\
 & 8 &    -63648.9139   &   -1515457.41   &    -- \\
 & 9 &     62158.0044  &     1223772.96  &    -- \\
 &10 &    -31778.3109   &   -554471.928 &    -- \\
 &11  &    6523.12231  &     107982.421 &    -- \\
\hline
\end{tabular}
\label{t4}
\end{table}}

\newpage

{
\begin{table}[h]
\caption{Parameters for the $\psi_a(2)$ component}

\begin{tabular}{|c|c|r|r|r|}
\hline
     $L$ & $i$ & $A_i$& $A_i'$ &  $B_i$   \\
     \hline
     \hline
2 &0  &0.0906145663   &  -54.7995054 &    -- \\
 & 1 & -0.712220945  &    256.049408   & -57.8195768\\
 &2  &  -20.3136219   &   470.222091  &   211.630253\\
 &3  &   81.4188182   &   -8294.9022   & -385.318964\\
 & 4 &  -544.595070   &   33870.9853  &   367.461861 \\
 &5  &   2355.01212   &  -77139.6959  &  -137.187405 \\
 &6  &  -6864.79033  &    118585.336 &    -- \\
 &7  &   13318.8325  &   -138016.983 1 &    -- \\
 & 8 &  -16303.3461  &    126552.418 &    -- \\
 &9  &   11929.7419   &  -84889.6419  &    -- \\
 &10  & -4708.14062   &   35162.7881  &    -- \\
 &11  &  756.740280  &   -6501.81818 &    -- \\
\hline
\end{tabular}
\label{t5}
\end{table}}

{
 \begin{table}[h]
\caption{Parameters for the $\psi_t^0(1)$ component}

\begin{tabular}{|c|c|r|r|r|}
\hline
     $L$ & $i$ & $A_i$& $A_i'$ &  $B_i$   \\
     \hline
     \hline
1 &0 &  0.0126177910  &   0.547674806  &    -- \\
  &1 & -0.0622969395   &  -3.75788455  &  0\\
  &2 &    1.64026310  &    30.8407337  &  0\\
  &3 &   -25.8788379  &   -220.882836  &  0\\
  &4 &    241.662795   &   1370.42460   &   7.63938144 \\
  &5 &   -1410.95190   &  -5648.39611   &  -31.1836535\\
  &6 &    5126.52517  &    15857.5305  &    24.3738723\\
  &7 &   -12077.1617  &   -30611.8247 &    -- \\
  &8 &    18231.6691  &    39147.3503  &    -- \\
  &9 &   -16701.2532   &  -31118.3495 &    -- \\
  &10 &   8364.33221   &   13774.6556  &    -- \\
  &11 &  -1750.41655  &   -2578.09116 &    -- \\
\hline
\end{tabular}
\label{t6}
\end{table}}

\newpage

{
 \begin{table}[h]
\caption{Parameters for the $\psi_t(0)$ component}

\begin{tabular}{|c|c|r|r|r|}
\hline
     $L$ & $i$ & $A_i$& $A_i'$ &  $B_i$   \\
     \hline
     \hline
0 & 0&-0.000764090007   & -0.909552636  &    -- \\
 & 1 &  0.00774111957  &    3.53114793  & 0.0266757851\\
 &2  &    0.081210854   &  -79.1547771  &   4.16947945\\
 & 3 &     8.49620518   &   1058.66511  &  -3.26685212\\
 & 4 &    -122.622299   &  -8772.66567  &  -36.1206906\\
 &5  &     1056.75816   &   43565.4328   &  36.1389858\\
 & 6 &    -5058.22423   &  -132777.961 &    -- \\
 & 7 &     15091.7728  &    258802.623  &    -- \\
 & 8 &    -28171.6682   &  -327964.927 &    -- \\
 & 9 &     31347.0373  &    263416.014  &    -- \\
 & 10 &   -18926.2139  &   -122064.787  &    -- \\
 &11 &     4783.25176   &   24814.8930  &    -- \\
\hline
\end{tabular}
\label{t7}
\end{table}}

{
 \begin{table}[h]
\caption{Parameters for the $\psi_t(2)$ component}

\begin{tabular}{|c|c|r|r|r|}
\hline
     $L$ & $i$ & $A_i$& $A_i'$ &  $B_i$   \\
     \hline
     \hline
2 & 0 &   0.0927963894  &   6.63409360 &    -- \\
 & 1 &    -0.566806578   &  27.0403716  &  -26.4480841\\
 & 2 &     -6.23310249  &   628.891454   &  39.0979395 \\
 & 3 &     -22.0317803   & -5172.06340  &  -12.4959784 \\
 & 4 &      299.860807   &  22767.0450  &    -- \\
 & 5 &     -2053.92122  &  -74280.5598 &    -- \\
 & 6 &      8124.81954  &   177690.047  &    -- \\
 & 7 &     -20191.0789   & -294391.655 &    -- \\
 &8  &      31928.1425   &  323661.633  &    -- \\
 & 9 &     -30700.6358  &  -224618.740 &    -- \\
 & 10 &     16220.4293  &   89089.3349 &    -- \\
 &11  &    -3598.88935  &  -15407.7777  &    -- \\
\hline
\end{tabular}
\label{t8}
\end{table}}

\newpage

\centerline{\large \bf Figure captions}

 \begin{figure}[h]
\caption{ The diagram for  the matrix element
of operator $\hat O$ over the deuteron state in
the  impulse
approximation. }
\label{tre}
\end{figure}
\phantom{.}

 \begin{figure}[h]
\caption{ Densities of the vector (solid line) and axial (dashed line)
 charges calculated with the presented
parametrization.}
\label{den}
\end{figure}

\phantom{.}

\newpage
 
 \phantom{.}
 
 \vskip 2cm

 \let\picnaturalsize=N
\def\picsize{10cm}
\vspace*{2cm}
\def\picfilename{tre.ps}
\ifx\nopictures Y\else{\ifx\epsfloaded Y\else\input epsf \fi
\let\epsfloaded=Y
\centerline{\ifx\picnaturalsize N\epsfxsize
 \picsize\fi \epsfbox{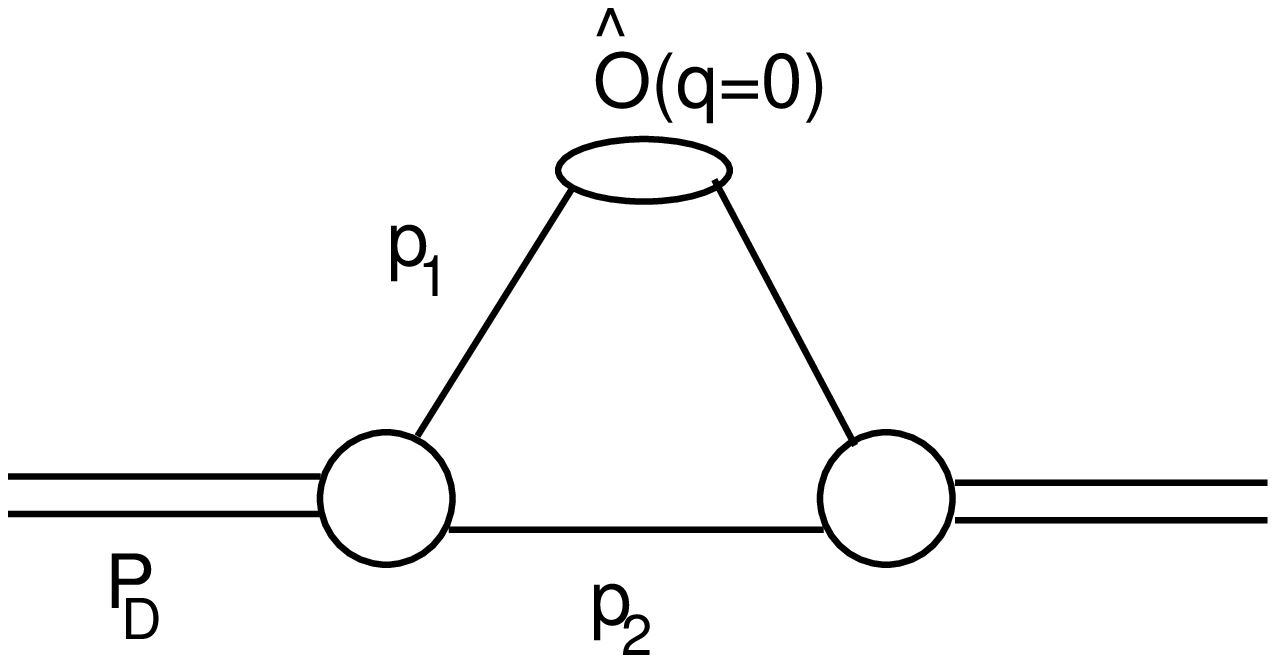}}}\fi
\vfill
Fig.~\ref{tre}. A. Umnikov...

\phantom{.}


 \let\picnaturalsize=N
\def\picsize{15cm}
\def\picfilename{amp-den.ps}
\ifx\nopictures Y\else{\ifx\epsfloaded Y\else\input epsf \fi
\let\epsfloaded=Y
\centerline{\ifx\picnaturalsize N\epsfxsize
 \picsize\fi \epsfbox{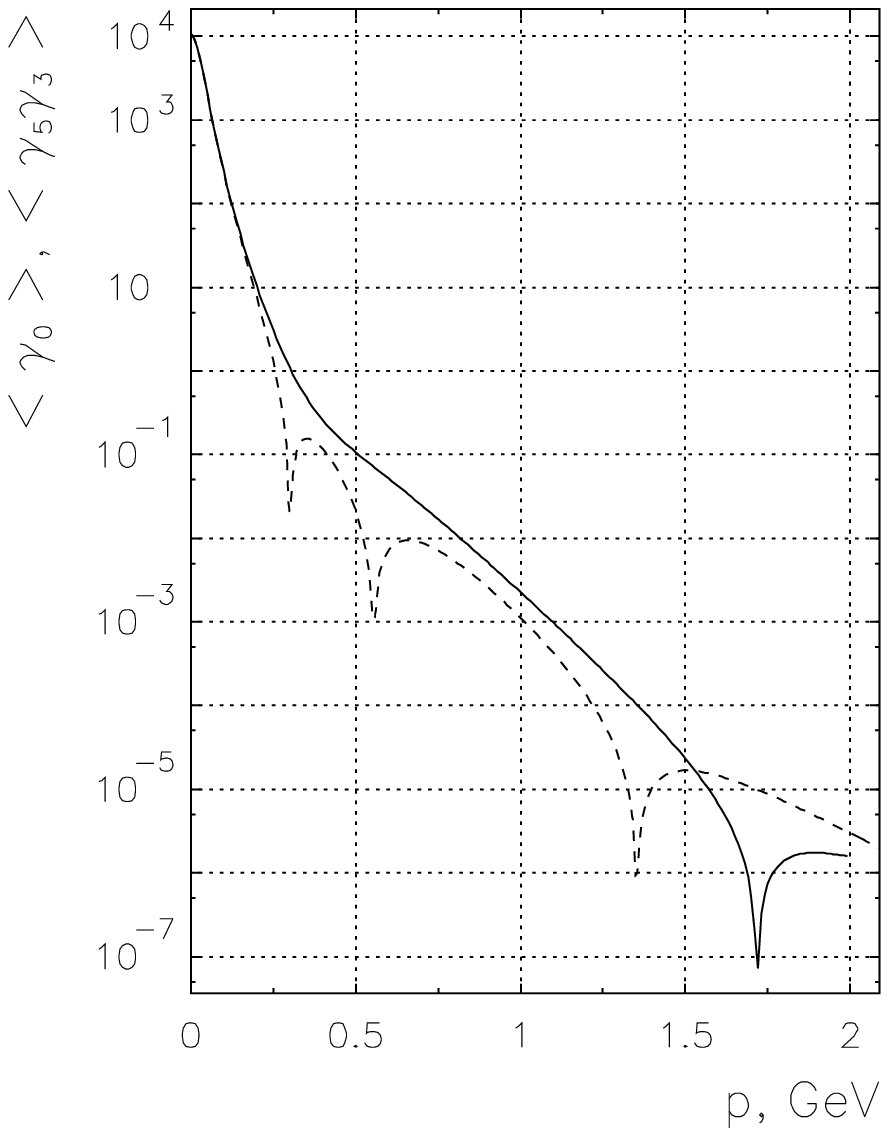}}}\fi
\vfill
Fig.~\ref{den}. A. Umnikov...


\begin{thebibliography}{50} 
  \bibitem{tjond}  M.J. Zuilhof and J.A. Tjon, Phys. Rev. {\bf C22} (1980) 2369.
  \bibitem{tjon2} B.D. Keister and J.A. Tjon, Phys. Rev. {\bf C26} (1982) 578.
\bibitem{ukkk} A.Yu.  Umnikov, L.P.  Kaptari, K.Yu. Kazakov
 and F. Khanna, Phys. Lett. {\bf B334} (1994) 163.
\bibitem{kuk} L.P.  Kaptari, A.Yu.  Umnikov,  B. K\"ampfer
 and F. Khanna, Phys. Lett. {\bf B351} (1995) 400.
\bibitem{ukk} A.Yu.  Umnikov, L.P.  Kaptari 
 and F. Khanna, Alberta-Thy-29-94; e-print archives hep-ph 9410241 and
 {\em in preparation}.
 \bibitem{paris} M. Lacombe et al., Phys. Rev. {\bf C21} (1980) 861.
\bibitem{bonn} R. Machleid, K. Holinde and Ch. Elster, Phys. Rep.
 {\bf 149} (1987) 1.
\bibitem{uk} A.Yu.  Umnikov 
 and F. Khanna, Phys. Rev. {\bf C49} (1994) 2311.  
 \bibitem{wolf} S. Wolfram, {\em Mathematica}, Addison-Wesley (Reading, Massachusetts, 1993).
 \bibitem{paris1} M. Lacombe et al., Phys. Lett. {\bf B101} (1981) 139.
 \bibitem{buckgross} W.W. Buck and F. Gross, Phys. Rev. {\bf C20} (1979) 2361.
 \bibitem{tjonnn} J. Fleischer and J.A. Tjon, Nucl. Phys. {\bf B84} (1975) 375;
                   Phys. Rev. {\bf D15} (1977) 2537.
\end{thebibliography}
\end{document}